\documentclass[11pt]{amsart}
\usepackage{geometry}                % See geometry.pdf to learn the layout options. There are lots.
\usepackage{graphicx}
\usepackage{amssymb}
\usepackage{epstopdf}
\usepackage{pgfplots}

\begin{document}

%\DeclareGraphicsRule{.tif}{png}{.png}{`convert #1 `dirname #1`/`basename #1 .tif`.png}

\title{Exploring the parameter space of disc shaped silver nanoparticles for thin film silicon photovoltaics}
\author{Bruno Figeys, Ounsi El Daif\\{IMEC vzw, Leuven, Belgium}
}
\date{}                                           % Activate to display a given date or no date

\maketitle

%%%%%%"%%%%%%%%%%%%%%
\begin{abstract}
We numerically simulate, using finite-difference time-domain, the optical properties of silver nano discs deposited on the front surface of silicon solar cells. We explore the effect of each of the parameters of such a system, in order to draw some general design rules for the subsequent fabrication of such structures.
\end{abstract}

\section{Introduction}
Placed on the interface of two media with a high and a low refractive index, metallic nanoparticles scatter a higher fraction of light into the high refractive index medium\footnote{This work is adapted from a Masters thesis done at the Katholieke Universiteit Leuven\cite{masters}}. In addition, the scattered light has a broad distribution of angles, increasing the lightÕs path length through this medium. These are the two effects that make silver nanoparticles promising for solar cells. In particular the increase in path length enables an increase in the absorption when the active region of the solar cell is small. However the introduction of metal particles will also introduce additional ohmic losses. The overall transmission of solar incident light into the high refractive index material, may therefore be smaller than in the case of an optimised antireflective coating\footnote{a single layer of 80 of silicon nitride is often used in silicon photovoltaics}. It has been reported as well that for particles placed at the sun-facing side of the cell, inside the high index material the incoming light interferes with the light scattered by the nanoparticles. For light energies above the plasmon resonance, where the electrons show a phase lag compared to the exciting light, this means that destructive (Fano) interferences occur, leading to an absorption suppression. Consequently, as the solar power is maximum in the green part of the spectrum, the plasmon resonance should be designed at higher energies, corresponding to blue wavelengths.

Both the fraction of scattered light into the solar cell as well as the increase in path length can be maximised by positioning the particle as close as possible to the high refractive index material. Because of this we consider in this work a silica ($SiO_{2}$) optical spacer of $\approx 10 - 20 nm$, and to ensure the transmission rivals that of an antireflective coating, a cladding of $\approx 100 nm$ of $SiO_{2}$ on top of the particles is considered. The ideal density of the particles depends on the inter-particle coupling. For too small inter-particle distances, large inter-particle coupling occurs, leading to a higher reflectivity and losses. For the simulations the periodicity is set to $300nm$, except when the effect of this inter-particle distance is specifically studied. The particle size also plays a crucial role: the particle resonance will red shift as the diameter of the particle increases.

We performed numerical simulations using Lumerical\cite{lumerical}, a finite-difference time-domain (FDTD) simulation software package. A plane wave source was simulated at the air side of an air/silicon interface with a $SiO_{2}$ interface layer, including a silver disc. The top and bottom boundary condition were perfectly matched layers. Only a quarter particle was simulated, the vertical boundary conditions were therefore chosen symmetric for boundaries parallel to the polarization of the incoming wave and anti-symmetric for boundaries perpendicular to the polarization. This way a particle array is simulated and the simulation time reduced. Using the transmitted power above and below the particle array and 400nm into the silicon, the absorbed power in the particle array, the absorbed power in silicon and the total transmitted and reflected power can be calculated. The absorbed power in silicon is a measurement of the scattering in different angles. Indeed, when the path length is increased it will cause an increase in absorption in this thin region. The dielectric functions for Ag as well as for Si were based on experimental values found in Palik's reference book\cite{palik}.

This article describes the trends of the dipolar surface plasmon resonance observed for changing parameters, such as the particle diameter or the spacer thickness. We considered as benchmark structures two different structures: a bare silicon layer without anything at the surface, and a silicon layer with 80nm of silicon nitride (Si$_3$N$_4$) as an antireflection coating, which is usually used in solar cells.

%%%%%%%%%%%%%%%%%%%%
\newpage
	\section{Simulation structure}
\begin{figure}[!htp]
\includegraphics[width=0.9\textwidth]{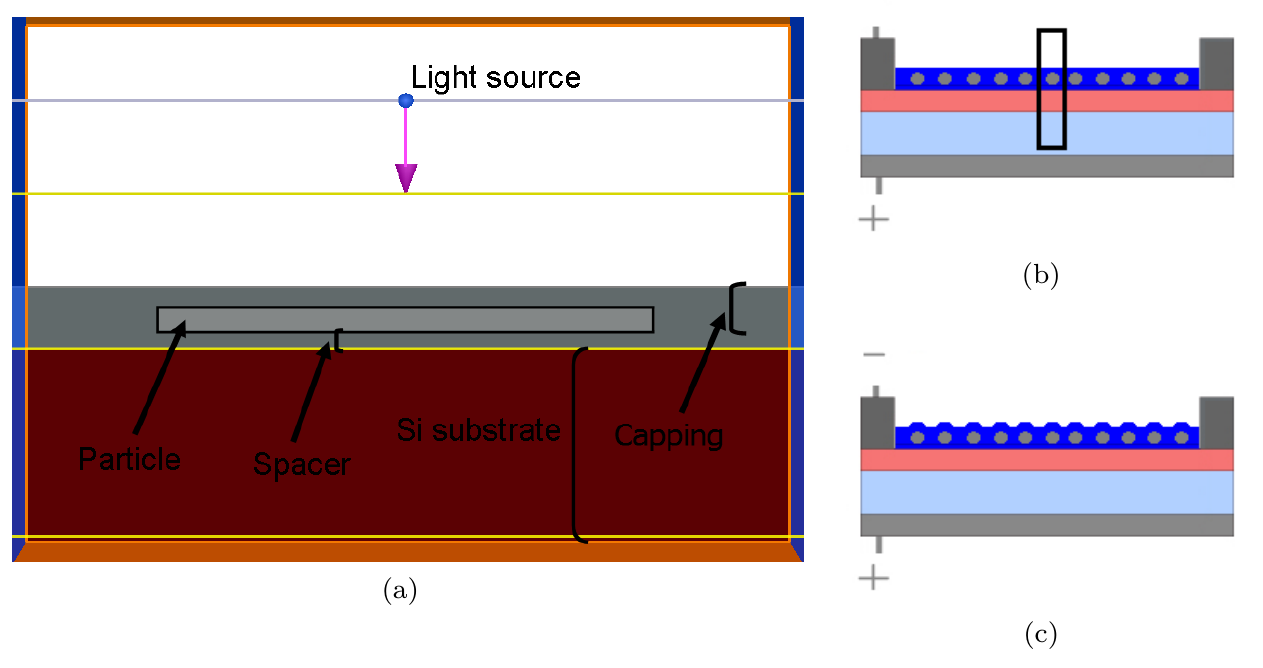}
\caption{Simplified drawing of a solar cell with nanoparticles embedded in an oxide. (b) The black box represents the part that is simulated.}
	\label{oneparticle}
\end{figure}

Figure \ref{oneparticle} shows the structure that is used in the simulations. It consists of a layer of silicon (brown), with on top of it a layer of silicon oxide (dark grey). In this layer of silicon oxide there is a silver disc (light grey). The particle is at a certain distance from the silicon layer, the thin layer between the silicon and the particle is the spacer, whereas the layer above the particle is the capping.

The main elements used in the present work are the following:
\begin{description}
\item[Silicon substrate] The silicon layer is the active layer of the solar cell, where light absorption should be enhanced. The electrical properties of the solar cell are not simulated with Lumerical, therefore only the number of absorbed photons is calculated. The silicon layer considered has been taken thin ($\approx 300-400nm$) to reduce the calculation time.

\item[Spacer] The spacer between the silicon and the particle is necessary for several purposes. 
\begin{enumerate}
\item In order to prevent material contamination in the silicon, as a barrier between silver and silicon of the metal into the silicon.
\item The optical spacer allows blue-shifting the surface plasmon resonance (SPR) wavelength. Previous works\cite{BeckPolmanCatchpole} indicate that it is better to stick to silicon oxide for particles on the frontside.
\end{enumerate}

\item[Particle] The particle, a disc, has two parameters that can be tuned: its radius and thickness. Silver has been used for its low ohmic losses as well as for its smaller detrimental effect on silicon properties, in case of contamination (compared e.g. to copper or gold). The reason for this comes from the mid bandgap energy levels copper and gold impurities generate in silicon band structure, as can be seen on figure \ref{impuritysilicon}. These mid bandgap energy levels degrade the carrier relaxation time strongly due to an increased Shockley-Read-Hall recombination.  Each recombination of an electron-hole pair represents a photon that has been absorbed in the silicon and that has not been converted into current. It thus reduces the IQE.

\begin{figure}[htbp]
	\centering
		\includegraphics[width=0.8\textwidth]{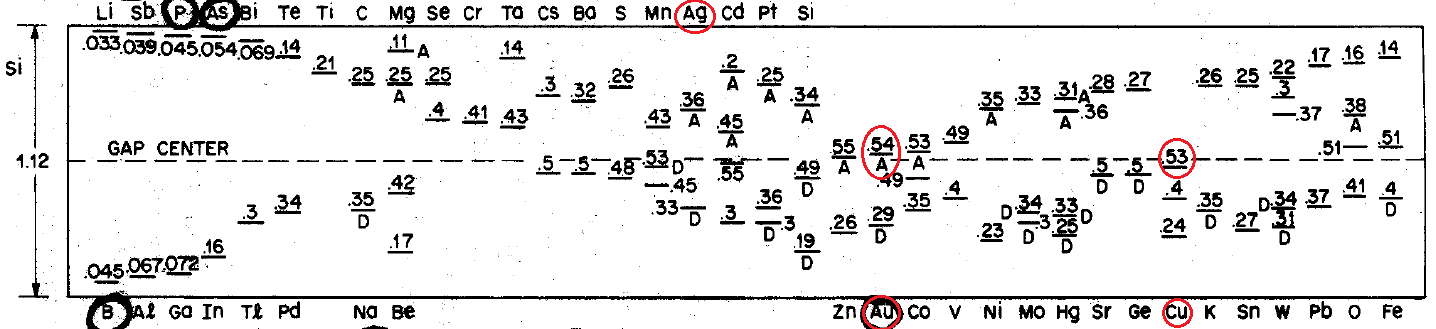}
	\caption{Representation of energy levels of impurities in silicon bandgap.}
	\label{impuritysilicon}
\end{figure}

\item[Capping] The capping which is the top part of the dielectric has also been chosen to be silicon oxide. This layer acts as an antireflection coating, complimentary to the platonic effect for off-resonance wavelengths. Experimentally it can also improves the stability of the nanoparticles by preventing their oxidation in air.

\end{description}

%%%%%%%%%%%%%%%%%%%%
%\newpage
	\section{Simulation setup}	
The simulation structure was made as small as possible, as the simulation time can increase rapidly due to the small mesh size necessary around the silver nanodiscs: in order to account for the plasmon resonance, which is mainly characterized by an enhancement electromagnetic field around the particle. It has therefore been chosen to do the simulations only for one particle instead of an array. For faster simulations it has been chosen to use periodic structures on the sidewalls, using the symmetric and anti-symmetric boundary conditions, as described higher in the introductory part.

%%%%%%%%%%%%%%%%%%%%
\newpage
\section{Diameter of the nanodisc}
\begin{figure}[htbp]
   \includegraphics[width=0.9\textwidth]{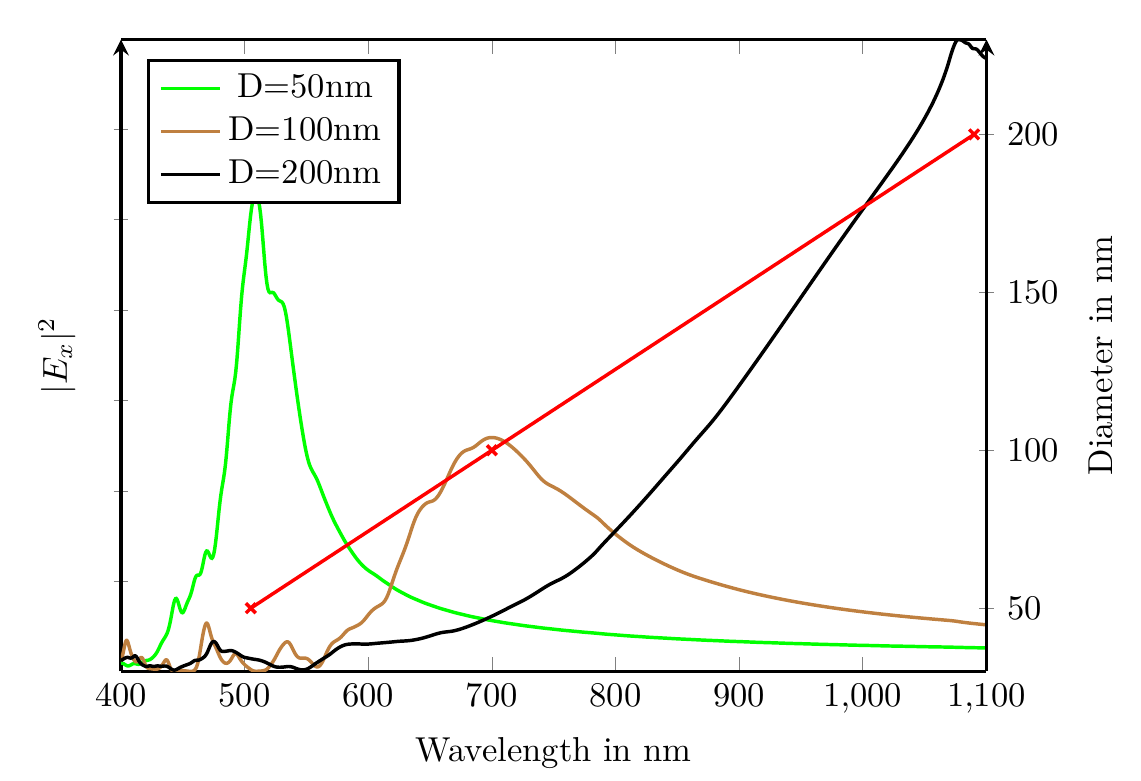}
	\caption{Influence of particle diameter on SPR}
	\label{Ediameter}
\end{figure}

\begin{figure}[htbp]
\includegraphics[width=0.9\textwidth]{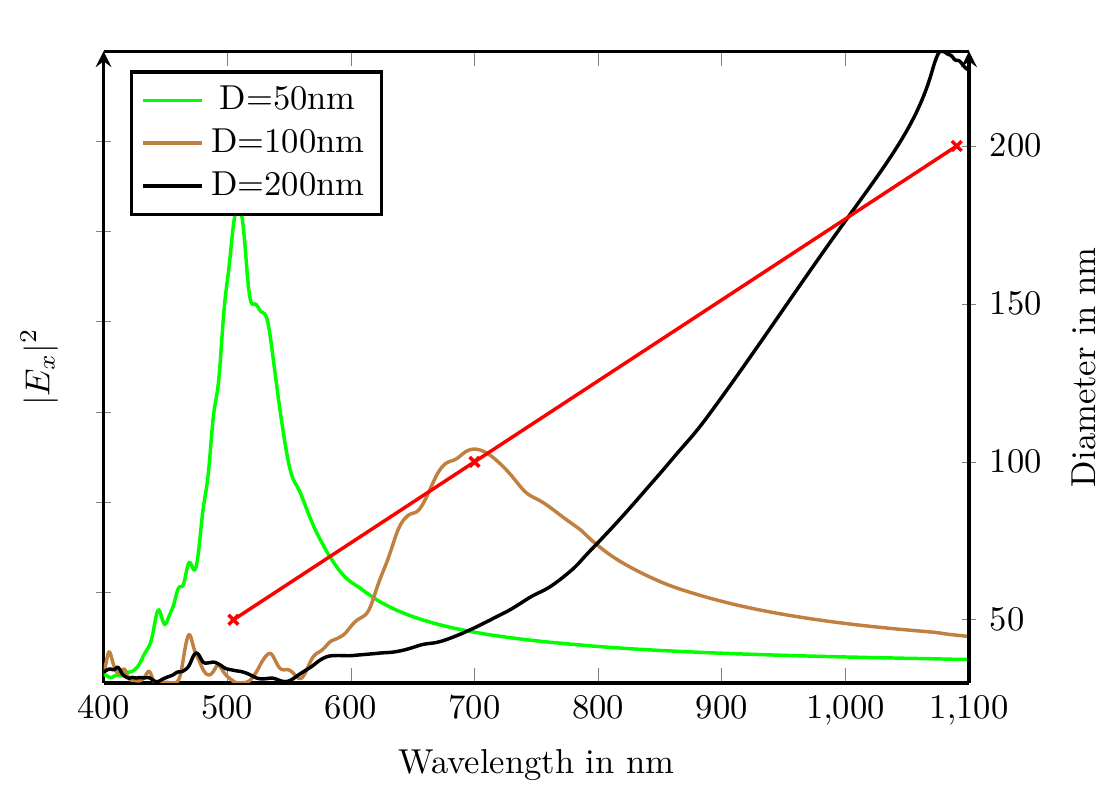}
	\caption{Influence of particle diameter on reflectivity, coupling into silicon and absorption in the particle.}
	\label{Rdiameter}
\end{figure}

As can be seen in figures \ref{Ediameter} the SPR red-shifts with an increasing disc diameter. This implies that the Fano resonance characteristic also red-shifts, as can be seen in figure \ref{Rdiameter}. The coupling of light into silicon shows a minimum around $625nm$ and $925nm$ for $100nm$ and $200nm$ diameter respectively. Considering the solar spectrum this is a detrimental spectrum as an important amount of light would be reflected at a wavelength where the solar spectrum is maximum. The bottom figure shows the absorption of light in the particle by ohmic losses. This is around 10\% for 100nm diameter particles but increase up to 20\% for 50nm diameters.

%%%%%%%%%%%%%%%%%%%%%%%%%%%%%%%
\newpage
\section{Thickness of the nanodisc}
\begin{figure}[htbp]
\includegraphics[width=0.9\textwidth]{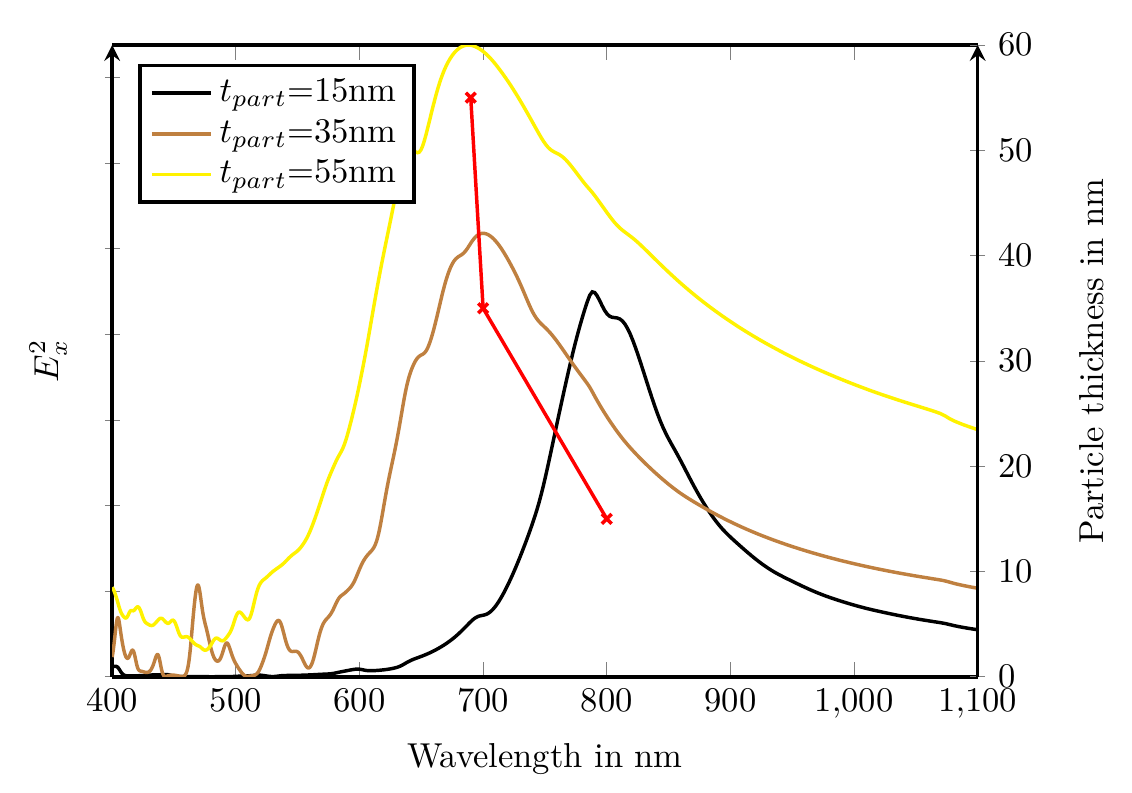}
	\caption{Influence of particle thickness on SPR.}
	\label{Ethick}
\end{figure}

\begin{figure}[htbp]
\includegraphics[width=0.9\textwidth]{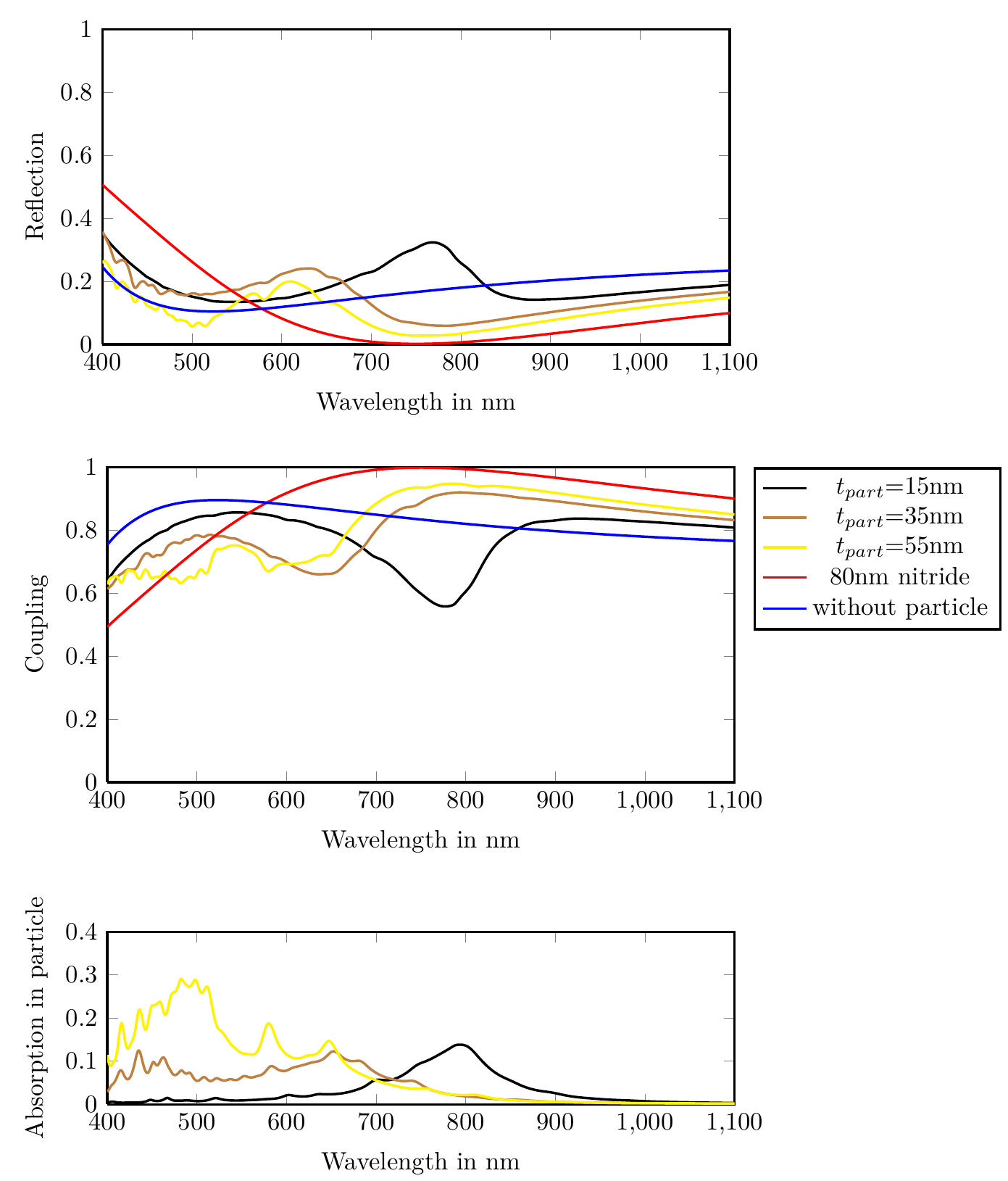}
	\caption{Influence of particle thickness on reflectivity, coupling into silicon and absorption in the particle.}
	\label{Rthick}
\end{figure}

Figure \ref{Ethick} shows that for thicker particles, i.e. towards cylinders, the SPR blue-shifts. The same figure also shows that a certain saturation is reached for a thickness of $55nm$. The same trend is naturally to be seen in the reflectivity graph in figure \ref{Rthick}.

%%%%%%%%%%%%%%%%%%%%%%%%%%%%%%%
\newpage
\section{Thickness of the spacer}
\begin{figure}[htbp]
	\includegraphics[width=0.9\textwidth]{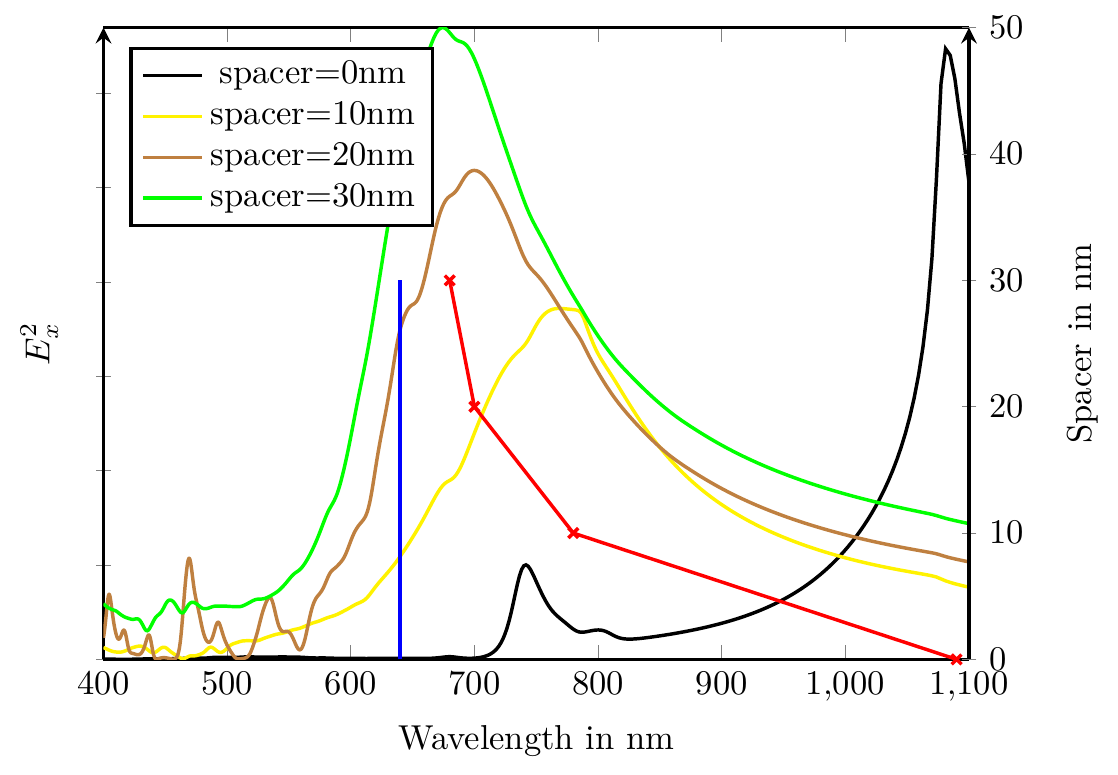}
	\caption{Influence of the spacer thickness on SPR. The vertical line represent the SPR in silicon oxide without the influence of the silicon.}
	\label{Espacer}
\end{figure}

\begin{figure}[htbp]
\includegraphics[width=0.9\textwidth]{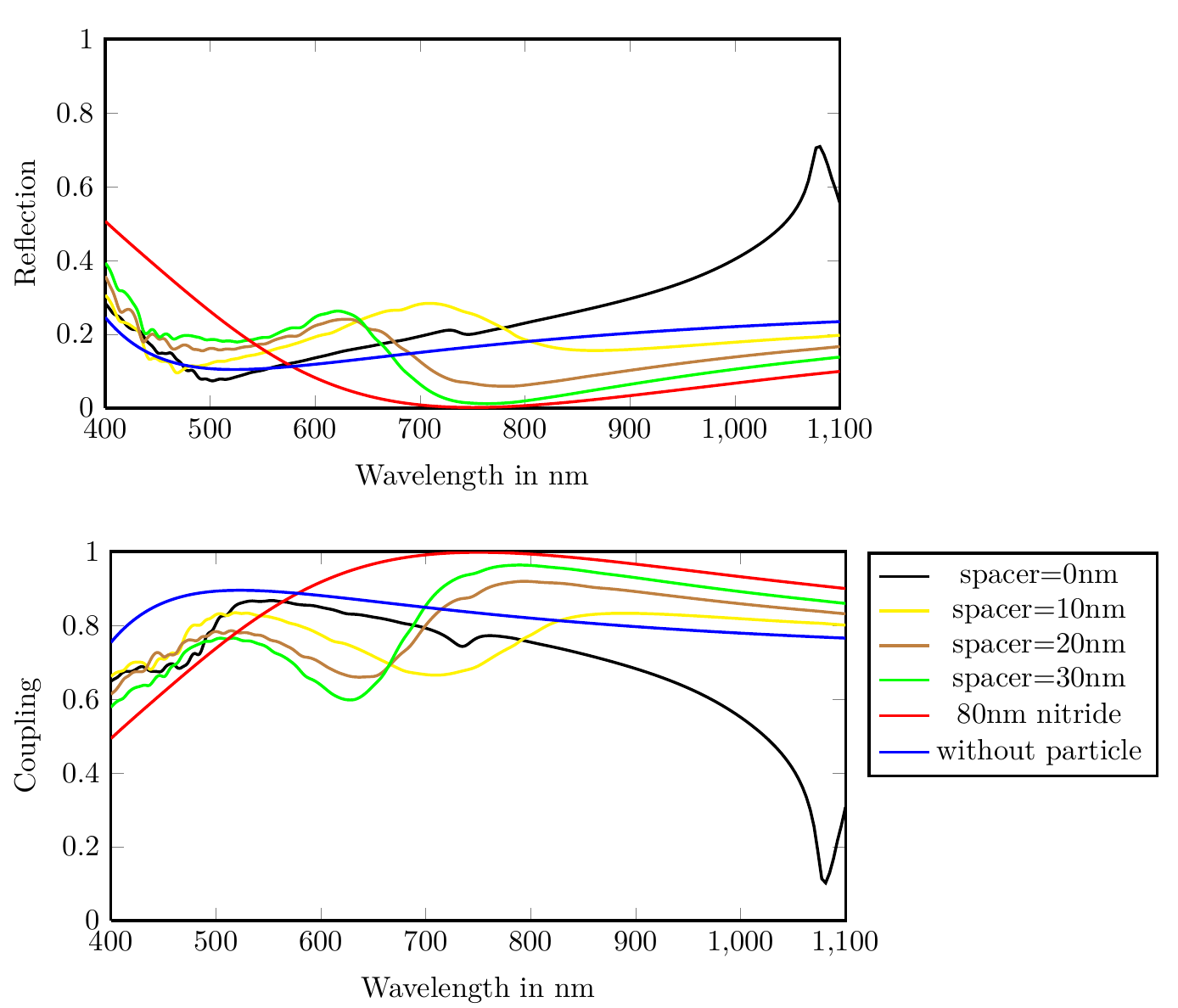}
	\caption{Influence of spacer thickness on reflectivity, coupling into silicon.}
	\label{Rspacer}
\end{figure}

Figures \ref{Espacer} and \ref{Rspacer} show that for a thinner spacer the resonance red-shifts. This is the same trend as for an increasing refractive index of the medium.  This suggests that by bringing the dipolar oscillator closer to the surface the average refractive index is increased, leading to a red-shift.

The resonance seems to go towards a limit for large spacer thicknesses. This limit is the SPR for the particle in an oxide environment without the influence of the silicon indicated with the blue line. A second observation is that the Fano-modulation around the curve without particle is stronger for a larger spacer. This is a confirmation of the proposed increase in scattering cross section due to a larger effective electric field further away from the substrate as has been proposed in the paper of K. Catchpole and A. Polman \cite{catchpol20082}.

\section{Periodicity of the particles}
\begin{figure}[htbp]
\includegraphics[width=0.9\textwidth]{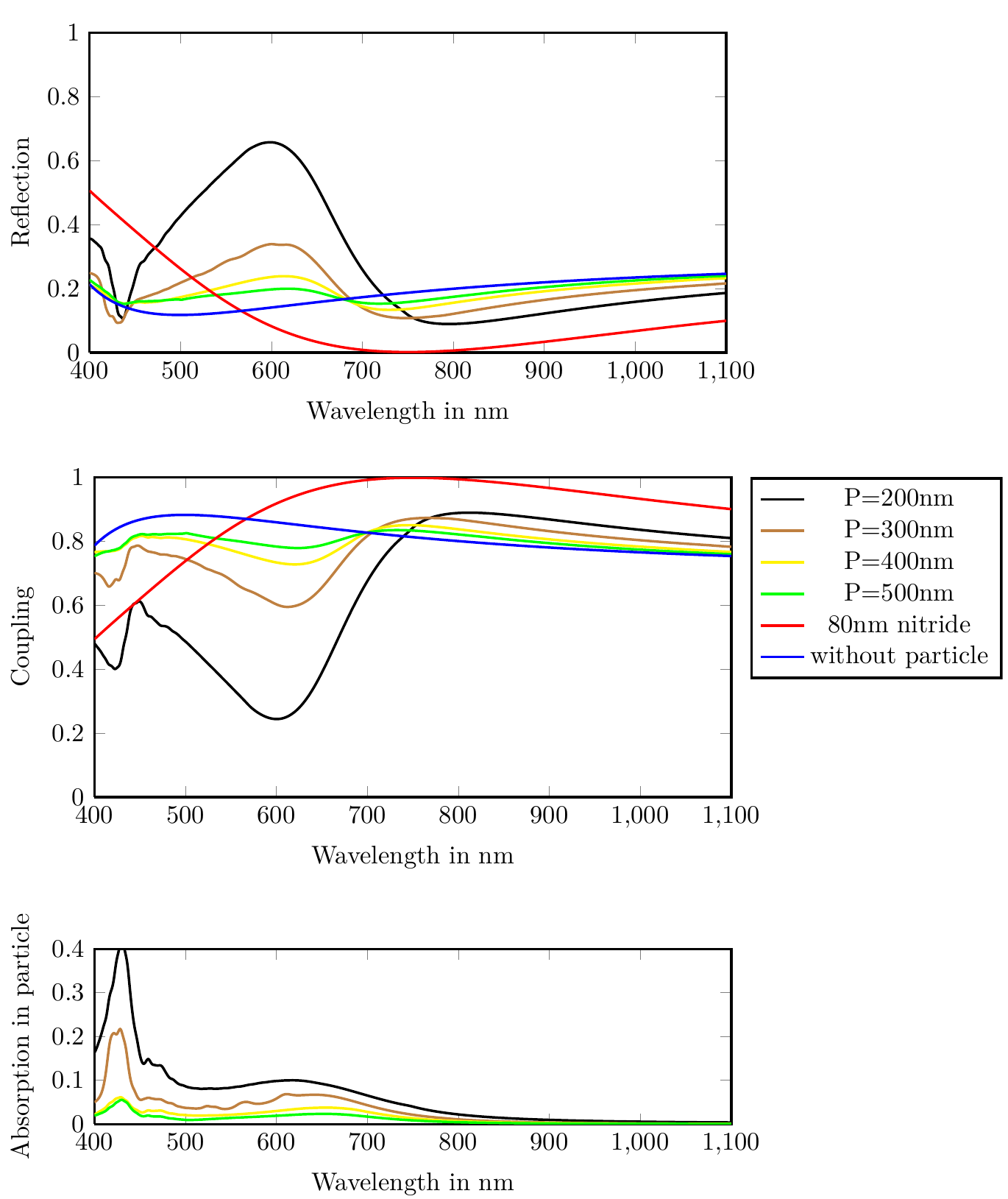}
	\caption{Influence of the periodicity of the particles on reflectivity, coupling into silicon and absorption in the particle. Capping is 50nm.}
	\label{RP}
\end{figure}

Large periodicities don't influence the resonance wavelength, whereas for small periodicities, high densities, it red-shifts. This is visible in figure \ref{RP}, for $P=300$ to $500nm$ all curves cross at the same point around $700nm$. For $P=200nm$ the resonance is red-shifted a bit. Figure \ref{RP} shows the Fano resonance consisting of a successive decrease and increase with respect to the curve of the structure without any particle. For smaller periodicities, thus higher density of particles, this Fano-modulation becomes larger and larger. This indicates that for too large periodicities the effect of coupling light into the silicon decreases because of a too low density of particles. The scattering cross section is thus too small and a lot of light is transmitted without being scattered. The absorption of the particle is the same, but as the density of particles increases for smaller $P$, the absorption in particles per surface area of the solar cell increases rapidly.

  \begin{figure}[htbp]
\includegraphics[width=0.9\textwidth]{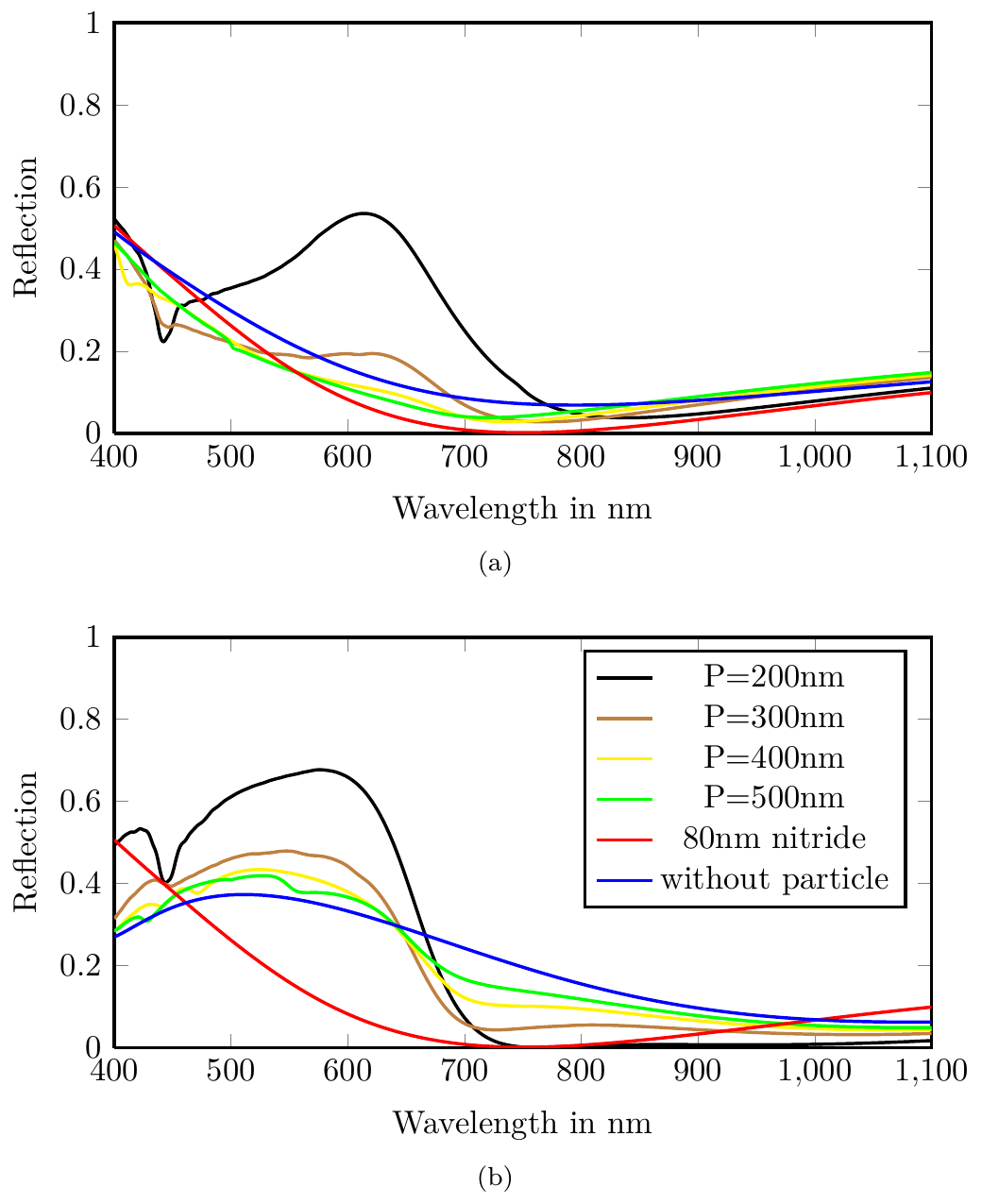}
	\caption{Influence of the periodicity of the particles on reflectivity, coupling into silicon and absorption in the particle. Same as figure \ref{RP} but with thickness of capping of (a) 100nm and(b) 150nm.}
	\label{RP2}
\end{figure}

%%%%%%%%%%%%%%%%%%%%%%%%%%%%%%%
\newpage
\section{Thickness of the capping}
\begin{figure}[htbp]
\includegraphics[width=0.9\textwidth]{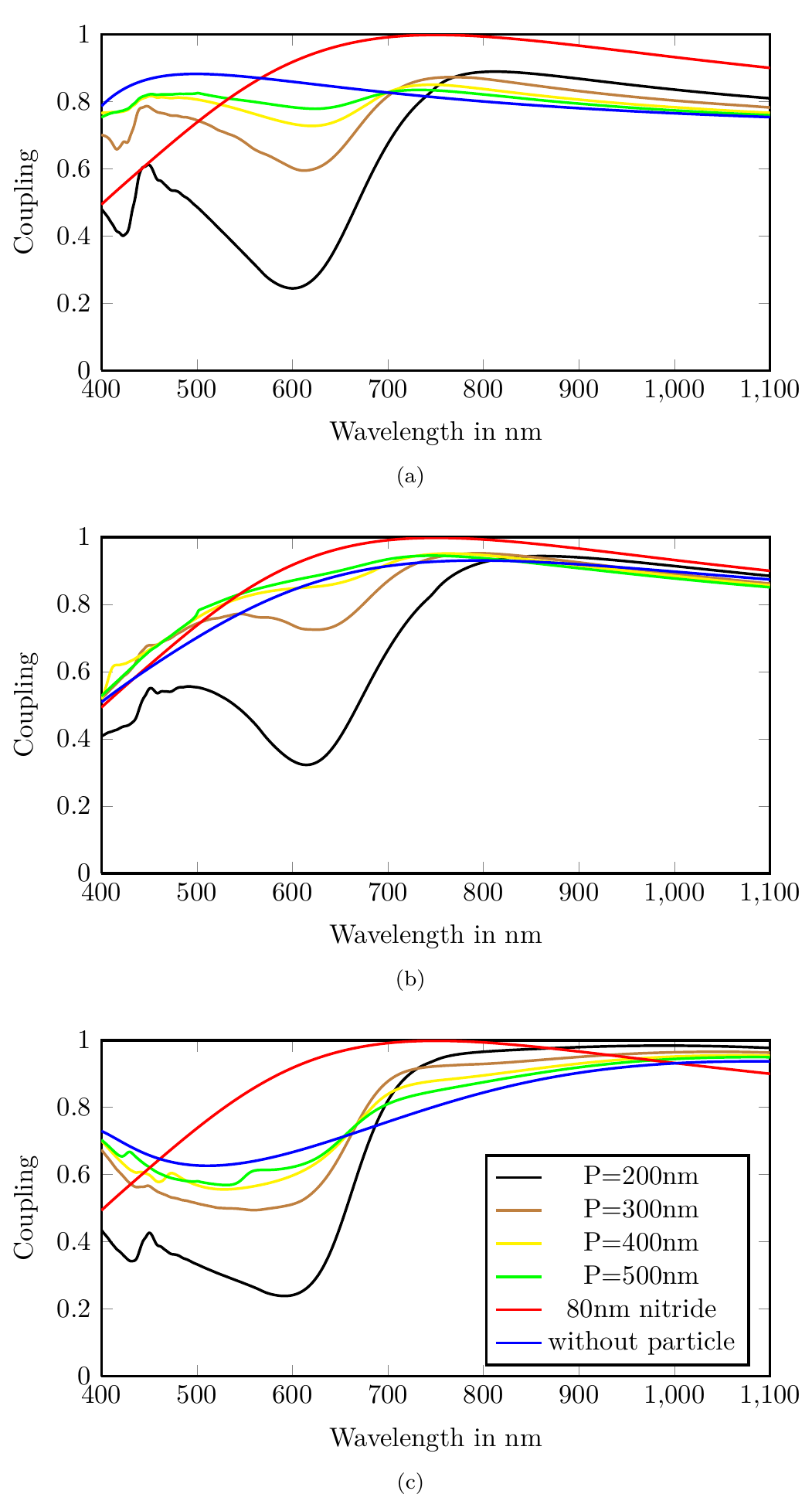}
	\caption{Influence of the periodicity and the capping thickness of the particles on the coupling into the silicon.  (a) 50nm, (b) 100nm and (c) 150nm oxide capping.}
	\label{Tclad}
\end{figure}

The thickness of the capping has no influence on the resonance wavelength. However figure \ref{Tclad} shows that the thickness of this layer of oxide for these large periodicities is important. The Fano-modulation can be seen as superimposed on the ARC effect of the oxide layer. The figures represented are for the same simulations as in figures \ref{RP} and \ref{RP2}.	

\bibliographystyle{abbrv}
\bibliography{references.bib}

\end{document}